\documentclass[conference, 9pt]{IEEEtran}
\IEEEoverridecommandlockouts
\usepackage{cite}
\usepackage{amsmath,amssymb,amsfonts}
\usepackage{bm}
\usepackage{algorithmic}
\usepackage{graphicx}
\usepackage{textcomp}
\usepackage{xcolor}
\usepackage{placeins}
\usepackage{colortbl}
\usepackage{hyperref}
\usepackage{tcolorbox}
\usepackage{booktabs}
\def\BibTeX{{\rm B\kern-.05em{\sc i\kern-.025em b}\kern-.08em
    T\kern-.1667em\lower.7ex\hbox{E}\kern-.125emX}}
\begin{document}

\title{Audio Codec Augmentation for Robust Collaborative Watermarking of Speech Synthesis
\thanks{This work was supported by JST, PRESTO Grant Number JPMJPR23P9, Japan. We acknowledge the computational resources provided by the Aalto Science-IT project and the TSUBAME4.0 supercomputer at Tokyo Institute of Technology. L.J.~used GitHub Copilot to assist in software development.}
}

%% TITLE CANDIDATES
% - Approximate Gradients for Audio Codecs Improve Collaborative Watermarking for Speech Synthesis
% - Audio Codec Augmentation for Robust Collaborative Watermarking of Speech Synthesis
% - Collaborative Watermarking through Audio Codecs

\author{\IEEEauthorblockN{%1\textsuperscript{st} 
Lauri Juvela}
\IEEEauthorblockA{\textit{DICE}, \textit{Aalto University}\\
Espoo, Finland \\
lauri.juvela@aalto.fi}
\and
\IEEEauthorblockN{%2\textsuperscript{nd}
Xin Wang}
\IEEEauthorblockA{\textit{National Institute of Informatics} \\
Tokyo, Japan \\
wangxin@nii.ac.jp}
% \and
% \and
% \and
% \and
}

\maketitle

\begin{abstract}
% Max 2000 characters

Automatic detection of synthetic speech is becoming increasingly important as current synthesis methods are both near indistinguishable from human speech and widely accessible to the public. Audio watermarking and other active disclosure methods of are attracting research activity, as they can complement traditional deepfake defenses based on passive detection. In both active and passive detection, robustness is of major interest. Traditional audio watermarks are particularly susceptible to removal attacks by audio codec application. Most generated speech and audio content released into the wild passes through an audio codec purely as a distribution method. We recently proposed collaborative watermarking as method for making generated speech more easily detectable over a noisy but differentiable transmission channel. This paper extends the channel augmentation to work with non-differentiable traditional audio codecs and neural audio codecs and evaluates transferability and effect of codec bitrate over various configurations. The results show that collaborative watermarking can be reliably augmented by black-box audio codecs using a waveform-domain straight-through-estimator for gradient approximation. Furthermore, that results show that channel augmentation with a neural audio codec transfers well to traditional codecs. Listening tests demonstrate collaborative watermarking incurs negligible perceptual degradation with high bitrate codecs or DAC at 8kbps.

% - Training time augmentation can help
% - Collaborative training requires gradients through codec, which is not trivial with black-box standard codecs
% - This paper shows that 
%     - Gradients can be reliably approximated with the STE
%     - Collaborative training improves robustness to codecs compared to observer (in low bitrates?)
%     - good generalization over different codec types?
%     - also works with neural audio codecs!
%     - 
%     - Listening tests show small perceptual trade-off with high codec bitrates and a gradual degradation at lower bitrates.
\end{abstract}


\section{Introduction}

% PAR 1
% instant voice cloning with TTS
% neural codec language models TTS
Modern text-to-speech (TTS) systems achieve close to human-level naturalness and are capable of zero-shot voice cloning from limited data \cite{jia2018-transfer-learning-from-speaker-verification}. Open-source implementations with pre-trained models have made voice cloning technology widely available to the research community \cite{casanova2022yourtts}, and hosted solutions for voice cloning are available to the public.  
More recently, neural audio codec language models have further improved the zero-shot voice cloning capabilities \cite{wang2023-vall-e-neural-codec-language-models}. The wide availability of voice cloning tools creates possibilities for misuse, which have classically been countered with deepfake detection methods.
To complement passive detection measures, generative model providers can apply active disclosure methods, such as watermarking.
Furthermore, the new European Union AI act \cite{EuroParl-EU-AI-Act} requires labeling AI generated content. Implementation of such labeling is still an open question, and special attention is needed to ensure these labels persist in when released into the wild.

% PAR 2
% Passive countermeasures
Research in audio deepfake detection has mostly focused on passive detection scenarios, such as anti-spoofing \cite{Wu2015} in automatic speaker verification (ASV). In this setting, the defender has no prior knowledge on incoming attack types and must train a detector machine learning model using a limited number of known attack types and training examples.
Results from recent detection challenges, including 
ASVspoof \cite{LiuASVspoof2021} and Audio Deep Detection (ADD) \cite{yiADD2022a}, demonstrate high detection performance for in-domain data with equal error rates (EERs) below 5\%. However, these models are susceptible to shortcut learning from spurious audio features \cite{LiuASVspoof2021, muller21_asvspoof, zhang21da_interspeech}, and have been shown to generalize poorly in mismatching data domains \cite{paul2017generalization, muller2022does}.

% PAR 3
% Active disclosure
% Watermarking
Continued research in passive detection remains important in adversarial scenarios, but makers of generative models should also contribute to safe and responsible use of generative AI by active disclosure of generated content.
Audio watermarking using deep learning methods is currently seeing increased research activity  \cite{pavlovic2022-robust-speech-watermarking-dnn, chen2023wavmark, oreilly2024maskmark, singh2024silentcipher, sanroman-2024-audioseal-proactive-detection-voice-cloning-watermarking}.
Audio watermarks are typically applied as post-processing to generated data. This pipeline approach works well in responsibly hosted systems, where public access is controlled via APIs. In contrast, open-source settings are essential for reproducible research\footnote{Code and demo samples for this paper are available at \url{https://github.com/ljuvela/collaborative-watermarking-with-codecs}
}, and the academics have a strong incentive for sharing code and pre-trained models to maximize research impact. However, non-integral watermarks can be trivial to remove in source code. Sharing code and pre-trained models is essential for reproducible research. 
Moreover, separate watermarks incur additional cost in processing, which is often non-negligible when using neural network watermarks.
%
% Integrated watermarking
% benefits of integrated watermarks:
% suitable for open-source settings
% no added runtime cost
% 
Integrated watermarking schemes for active disclosure have been recently posed for image synthesis
\cite{yu2022-responsible-disclosure-fingerprinting}, as well as neural vocoding \cite{juvela2024-collaborative-watermarking}.

% PAR 4
% Robustness to codecs
Robustness is important for both passive and active detection of generated content. 
Virtually all audio shared on the internet passes through a lossy compression, and this makes audio codecs perhaps the most prevalent and casually applied attack against watermarks and deepfake detectors.
% Perhaps the most common and casually applied attack against watermarks and deepfake detectors is lossy compression algorithms, since virtually all audio shared on the internet passes through a codec.
%
In anti-spoofing countermeasures, zero-shot generalization to codecs has room for improvement: EERs on speech passed through various codecs ranges between 15-30\% \cite{Liu2023-ASVspoof-deepfake-speech-detection-in-the-wild}.
A potential remedy is to augment the detector training by including decoded samples in the training data. This is straightforward to implement in passive detector settings, but an active watermark encoder-detector setting requires gradient flow between the models and codecs are generally non-differentiable.

%In a passive detector setting this only requires a standard implementation of a codec, since the system does not need to pass gradients through the augmentation. However, in an active encoder-detector setting, gradients are needed to adjust the watermark encoder to improve detectability. 

%Commonly used audio codecs are not differentiable (or at least differentiable implementations are not readily available)
%Training passive detector models does not require passing gradients to the synthesis model, and this setting can be easily augmented with non-differentiable perturbations, such as audio codecs. 
%However, adjusting a generative model to actively aid the detector requires gradients

\begin{figure*}[t!]
    \centering
    \includegraphics[width=0.8\linewidth]{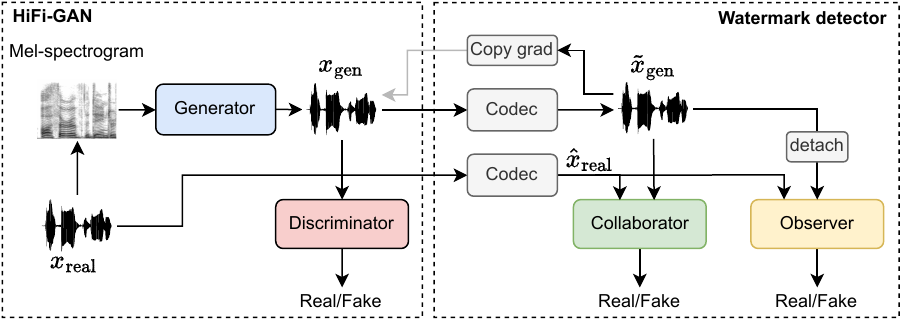}
    \caption{Generated speech can be detected by either a passive Observer or an active Collaborator. Watermark detection is made more robust against codecs by approximately differentiable channel augmentation required for passing gradients to the generator in collaborative training.
    }
    \label{fig:system-diagram}
\end{figure*}
% \begin{figure}[!ht]
%     \centering
%     \includegraphics[width=1.0\linewidth]{figures/system-diagram-v3.pdf}
%     \caption{System diagram}
%     \label{fig:system-diagram}
% \end{figure}

% Contributions:
This paper extends our previous work on collaborative watermarking \cite{juvela2024-collaborative-watermarking} of neural vocoders to use channel augmentation with non-differentiable audio codecs and neural audio codecs. %and compares transferability and effect of bitrate over various configurations.
The paper has the following contributions:
\begin{itemize}
    \item Show that codec augmentation improves robustness for both passive observer and active collaborator detector training.
    \item Confirm that the straight-through estimator works for gradient approximation in collaborative watermark training.
    \item Demonstrate that augmentation with a neural audio codec transfers well to robustness against traditional codecs. 
\end{itemize}

%%
%Similar results also extend to using the DAC neural codec 

\section{Related work}

Watermarking methods have various applications, and the specific application guides what trade-offs should be made with robustness, perceptibility, and capacity \cite{peticolas2000-watermarking-schemes-evaluation}.
For disclosing generated content, robustness is highly important, and watermark should be imperceptible enough not to hinder the primary use of generated content. However, useful capacity can potentially be as low as zero bits for a binary real/fake detection \cite{juvela2024-collaborative-watermarking}, i.e., the watermark is present if a sample is detected as fake.

% Classic audio watermarks
Classic signal processing-based audio watermarking techniques include
spread spectrum \cite{cox1996secure}, phase coding \cite{bender1996techniques},
echo hiding \cite{gruhl1996echo}, 
spectral patchwork statistics \cite{steinebach2004digitale}. However,
DSP watermarks are susceptible to removal with audio codecs, since watermarks often use similar perceptual models for information hiding as codecs use to discard information.

Robustness to codecs is of continued interest in neural audio watermarking methods. 
% high bitrates on codecs
While earlier work did not include codecs in their robustness evaluation
\cite{pavlovic2020-speech-watermarking-dnn, pavlovic2022-robust-speech-watermarking-dnn}, 
WavMark \cite{chen2023wavmark} includes one specific codec configuration: the system is quite robust to MP3 coding at 64kbps, with bit error rates between 0.0\% and 2.43\% on various datasets. Meanwhile, MaskMark \cite{oreilly2024maskmark} demonstrated high robustness to a traditional speech codec (OGG-Vorbis, unspecified at bitrate or quality level), while leaving room for improvement with neural codecs EnCodec \cite{dfossez2023-encodec-high-fidelity-neural-audio-compression} and DAC \cite{kumar2024-dac-high-fidelity-audio-compression}.
% MaskMark metric: True positive rate at 1% False positive rate
More recently, SilentCipher \cite{singh2024silentcipher} proposed using gradient copying to create a pseudo-differentiable codec for training time augmentation. This achieved both high robustness and imperceptibility for MP3, OGG (which is a container format, no codec was specified), and ACC, at bitrates 64kbps, 128kpbs, and 256kbps.

%New paper (looks like icassp submission) \cite{Roman2024-latent-watermarking-audio-generative-models}

% Benchmarking watermark robustness under codecs
% \cite{liu2024audiomarkbench}

% % Gradient approximation through non-differentiable operations
% \cite{bengio2013-ste-estimating-gradients-through-stochastic-neurons}
% \cite{van-den-oord-2017-neural-discrete-representation-learning}

% \cite{Zeghidour2021-SoundStream}
% \cite{kumar2024-dac-high-fidelity-audio-compression}

% Similar work soon published in IEEE SPL: 
% HiFi-GANw: Watermarked Speech Synthesis via Fine-tuning of HiFi-GAN (Cheng et al.)

% We are specifically interested in binary detection of generated content. This constitutes a zero-bit watermark. 
% Higher payloads are possible.

\section{Method}

The overall framework in this paper is similar to the collaborative watermarking scheme proposed earlier in \cite{juvela2024-collaborative-watermarking}. The main differences are introducing black-box DSP and differentiable neural codecs as the channel augmentation (detailed in section \ref{sec:gradient-approx-for-codecs} and updating the watermark detector model (section \ref{sec:models}).

\subsection{Collaborative training}
\label{sec:collaborative-training}

System overview for collaborative watermark training is shown in Fig. \ref{fig:system-diagram}.
Based on HiFi-GAN \cite{kong2020-hifi-gan}, 
a Generator network attempts to generate synthetic speech, given a mel-spectrogram as input.   
Other detector networks attempt to distinguish between the real samples and generator output.  
Depending on the Generators objective relative to the detector, the detector takes one of three roles

\begin{enumerate}
    \item[1)] \textbf{Discriminator} attempts to classify between real and generated waveforms, while the Generator attempts to fool the discriminator. Discriminator and Generator are adversaries.
\end{enumerate}
Discriminator training follows the typical generative adversarial network (GAN) \cite{goodfellow2014-generative-adversarial-nets} setting. %In this paper, we keep the original HiF
Watermark detectors also attempt similar classification, and are introduced as a third player in one of the two following roles:
\begin{enumerate}
    \item[2)] \textbf{Observer} attempts to passively classify between real and generated waveforms. However, no gradients are passed to the generator and the Generator is agnostic to the observer. This corresponds to classic anti-spoofing countermeasure training.
    \item[3)] \textbf{Collaborator} attempts to classify between real and generated samples, and the generator is actively trying to help the detector.
\end{enumerate}

Previous work \cite{juvela2024-collaborative-watermarking} used baseline models from the ASVspoof 2021 challenge \cite{LiuASVspoof2021} as detectors. In this paper, we upgrade the detector model to AASIST \cite{tak2022-aasist}.

\subsection{Gradient approximation for codecs}
\label{sec:gradient-approx-for-codecs}

Neural network training requires non-zero gradient flow through the system, but some operations disrupt the gradient and require special treatment.
Typical examples of such operations are related to learning vector quantized codebooks in discrete representation learning \cite{van-den-oord-2017-neural-discrete-representation-learning} and neural speech and audio codecs
\cite{Zeghidour2021-SoundStream, dfossez2023-encodec-high-fidelity-neural-audio-compression, kumar2024-dac-high-fidelity-audio-compression}, and audio codec gradient approximation \cite{singh2024silentcipher}.

% % Gradient approximation through non-differentiable operations

Copying gradients is implemented with the straight-through estimator \cite{bengio2013-ste-estimating-gradients-through-stochastic-neurons} using the following expression:
\begin{equation}
\label{eq:ste}
    \tilde{x} = \hat{x} + x - \lfloor x \rceil,
\end{equation}
where $x$ is the raw waveform, $\hat{x}$ is the decoded signal waveform, and $\lfloor \cdot \rceil$ is a detach operation that stops the gradient flow for backward pass. On forward pass, $x$ and its detached copy $\lfloor x \rceil$ cancel out numerically, resulting in only $\hat{x}$ passing forward. On backward pass, the expression in Eq.~\ref{eq:ste} effectively copies the gradient from $\tilde{x}$ to $x$:
\begin{equation}
    \frac{\partial\mathcal{L}}{\partial \tilde{x}} = \frac{\partial\mathcal{L}}{\partial \hat{x}}  + \frac{\partial\mathcal{L}}{\partial x} +  \frac{\partial\mathcal{L}}{\partial \lfloor x \rceil} =  \frac{\partial\mathcal{L}}{\partial x} ,
\end{equation}
because gradients are zeros for the non-differentiable terms $\partial \mathcal{L} / \partial \hat{x}$ and $\partial\mathcal{L} / \partial \lfloor x \rceil$.

\section{Experiments}

\subsection{Dataset}
All experiments in this paper use the LibriTTS-R \cite{koizumi23-libriTTS-R} dataset. LibriTTS-R is a large-scale multi-speaker English dataset derived from openly available audiobooks and uses audio restoration techniques to ensure a consistent audio quality over the full dataset. 
We used the \texttt{train-clean-100} (53.70\,h, 247 speakers) partition for training, \texttt{dev-clean} (8.97h, 40 speakers) for validation and the \texttt{test-clean} (8.56\,h, 39 speakers) partition for evaluation. The native sample rate for the data set is 24\,kHz, but the data was resampled to 22.05\,kHz for compatibility with pre-trained HiFi-GAN~\cite{kong2020-hifi-gan}.

\subsection{Models}
\label{sec:models}

We use HiFi-GAN as the neural vocoder for the experiments.
We adapted the official implementation  \cite{kong2020-hifi-gan}%\footnote{\url{https://github.com/jik876/hifi-gan}} 
and start fine-tuning from the public pre-trained V1 Universal  model. 
HiFi-GAN Generator is a fully convolutional 1D-ResNet that uses progressive upsampling to map from mel-spectrograms to speech waveforms. The Discriminator in HiFi-GAN consists of an ensemble of sub-discriminators: multi-period discriminators (MPD) operate on framed waveforms and use 2D convolution architecture, while multi-scale discriminators (MSD) use 1D convolution and progressively downsample the signal with strides and average pooling. All HiFi-GAN models operate at 22.05\,kHz sample rate. 

For a watermark detector, we use AASIST \cite{tak2022-aasist}. This model uses integrated spectro-temporal graph attention networks for audio anti-spoofing and is representative of state-of-the art in anti-spoofing. 
Likewise, we adapt the official implementation%\footnote{\url{https://github.com/clovaai/aasist/}} 
and use the provided pre-trained model as a starting point. The model operates at 16\,kHz sample rate, and we use the differentiable re-sampling in TorchAudio \cite{hwang2023torchaudio} to convert from the base sample rate of 22.05\,kHz.
Similarly to previous work on collaborative watermarking \cite{juvela2024-collaborative-watermarking}, dropout and batch normalization layers were removed from the detector. Collaborative training is related to adversarial training, and removing batch norm has been found to boost adversarial training \cite{wang2022-removing-batch-norm-boosts-adversarial-training}.

\subsection{Codecs for channel augmentation}
\label{sec:codecs-for-augmentation}

For channel augmentation experiments we selected two commonly used audio codecs, MP3 \cite{brandenburg1999-mp3-and-aac-explained} and Opus \cite{valin2013-opus-codec}. 
To quantify the effect of codec bitrates, we run the codecs at 16, 32, 64 and 128kbps. 
Augmentation chooses one codec and bitrate randomly from the pool of available options and the same configuration is applied to each minibatch element.
% actual training settings
Codec configurations used in training are listed in Table
\ref{tab:eer} with the results. 
All codecs run at 22.05\,kHz sample rate. %native to HiFi-GAN V1.
%
% Opus supports automatic mode detection for speech or audio coding modes, and we let the codec apply automatic detection.  
%
Additionally, we reserve the Vorbis codec %\cite{??}
as an unseen codec for evaluation.
The Vorbis encoder in 
FFMPEG does not support constant bitrate encoding, and we use quality scales 1, 2, and 3 in the evaluation. %Variable bitrate target bitrates by quality factors q=-2 for 32\,kbps, q=0 for 64\,kbps, and q=4 for 128\,kbps. 
%\cite{hydrogen-audio-vorbis-bitrates}
%
Opus and Vorbis use OGG as the container format, while MP3 uses its native container.

We implement codec augmentation by wrapping TorchAudio bindings of the FFMPEG library.
We opt to use the codecs with out-of-the-box support  for reproducibility and portability (either open or patents expired). Using a wider selection of speech codecs is feasible \cite{Liu2023-ASVspoof-deepfake-speech-detection-in-the-wild}, but this require custom FFMPEG builds with external libraries and licensing.
We release our implementation of pseudo-differentiable codec augmentation as part of a new toolkit\footnote{\url{https://github.com/ljuvela/DAREA}}, differentiable augmentation and robustness evaluation (DAREA).

For channel augmentation with a neural audio codec, we use the official pre-trained DAC \cite{kumar2024-dac-high-fidelity-audio-compression} model at 22.05\,kHz sample rate and 8\,kbps bitrate. DAC uses STE internally to approximate gradients through codebook quantization.

\subsection{Loss functions}

Generator and Discriminator loss functions are adopted from HiFi-GAN \cite{kong2020-hifi-gan} following the collaborative training described in \cite{juvela2024-collaborative-watermarking}. 
Denote generated speech signal as $x_\text{gen} = G(\text{MS}(x_\text{real}))$, where $G$ is the Generator model, and $\text{MS}(x_\text{real})$ is the mel-spectrogram of the target waveform.
Discriminator, $D$, is trained with the least-squares GAN loss function, with the target score at one for real samples and at zero for generated
\begin{equation}
    \mathcal{L}_{D} = \mathbb{E} \left[(D(x_\text{real})- 1)^2 + D(x_\text{gen})^2 \right], 
\end{equation}
where the expectation, $\mathbb{E}$, is approximated by minibatch averages over batch elements and timesteps. %Each of the sub-discriminators in the HiFi-GAN discriminator ensemble share the same objective. 
Similarly, the Generator loss functions are from HiFi-GAN, comprising the adversarial loss
% Adversarial, L1 melspec, feature matching
\begin{equation}
    \mathcal{L}_\text{G-adv} = \mathbb{E} \left[ \left( D(x_\text{gen} ) - 1 \right)^2 \right], 
\end{equation}
feature matching loss at each hidden activation $D_{i}(\cdot)$ in each sub-discriminator
\begin{equation}
    \mathcal{L}_\text{G-FM} =  \sum_{i} \mathbb{E} \left[ \left(D_{i}(x_\text{real})- D_{i}(x_\text{gen}) \right)^2 \right], 
\end{equation}
and L1 regression loss on log-mel-spectrograms
\begin{equation}
    \mathcal{L}_\text{G-mel} = \mathbb{E} \left[ |  \log (\text{MS} (x_\text{real}) ) -  \log ( \text{MS} (x_\text{gen}) ) | \right].
\end{equation}

% \begin{equation}
%     \mathcal{L}_\text{G-mel} = \mathbb{E} \left[ |  \log (\bm{M} \cdot | \text{STFT} (x_\text{real}) |) -  \log ( \bm{M} \cdot | \text{STFT} (x_\text{gen}) |) \ \right], 
% \end{equation}
% where $\bm{M}$ is a tensorized mel-filterbank matrix. 

\newcommand\WM{\mathit{WM}}

The watermark detector model, $\WM$, has a similar objective to $D$: assign a high score to codec-augmented $\hat{x}_\text{real}$ and a low score to codec-augmented and gradient-copying $\tilde{x}_\text{gen}$
\begin{equation}
    \mathcal{L}_{\WM} = \mathbb{E} \left[(\WM(\hat{x}_\text{real} )- 1)^2 + \WM(\tilde{x}_\text{gen})^2 \right]. 
\end{equation}
$G$ can either share $\mathcal{L}_{\WM}$ or ignore it by detaching gradients. These two scenarios are called Collaborator and Observer, respectively. 

% In practice, the system is trained by alternating minibatches that switch between the $D$ objectives and the joint objective of $G$ and $\WM$. In Observer mode, the gradient flow from $\WM$ to $G$ is detached and $\mathcal{L}_{\WM}$ has no effect on $G$.

\subsection{Training details}

% Key:
% 110: collaborator, no augmentation
% 129: observer,  mp3 and opus at 16
% 134: collaborator, dac at 8
% 135: observer, dac at 8
% 130, collaborator, mp3 and opus at 16
% 137: observer, no augmentation
% 131: observer,  mp3 and opus at 8
% 132: collaborator, mp3 and opus at 8

% Moved the result inclusion here to prevent table from going to page 5 
\begin{table*}[!t]
\caption{EER (\%) of experiment systems. A darker cell color indicates a higher EER value. Letters `o' and `c' denote observer and collaborator training modes, respectively. Bitrate (kpbs) is shown for codecs using a fixed bitrate, and quality scale (qscale) is shown for codecs with variable bitrates. }
\label{tab:eer}
\centering
\resizebox{1.0\linewidth}{!}
{
\resizebox{\textwidth}{!}{
\begin{tabular}{llcccccccccccccc}
\toprule
codec & & \multicolumn{2}{c}{None} & \multicolumn{2}{c}{DAC} & \multicolumn{10}{c}{MP3/Opus}   \\
\cmidrule(lr){3-4}\cmidrule(lr){5-6}\cmidrule(lr){7-16}
& bitrate & \multicolumn{2}{c}{-} & \multicolumn{2}{c}{8} & \multicolumn{2}{c}{16}  & \multicolumn{2}{c}{32}  & \multicolumn{2}{c}{64}  & \multicolumn{2}{c}{128}  & \multicolumn{2}{c}{all} \\
\cmidrule(lr){3-4}\cmidrule(lr){5-6}\cmidrule(lr){7-8}\cmidrule(lr){9-10}\cmidrule(lr){11-12}\cmidrule(lr){13-14}\cmidrule(lr){15-16}
                       &  &  o  &  c  &  o  &  c &  o  &  c &  o  &  c &  o  &  c &  o  &  c &  o  &  c  \\ 
                       %&  &  137  &  110  &  135  &  134  &  149  &  130  &  147  &  128  &  145  &  126  &  143  &  124  &  141  &  122 \\ 
\midrule
       None & - & \cellcolor[rgb]{0.99, 0.99, 0.99} 1.55 & \cellcolor[rgb]{1.00, 1.00, 1.00} 0.29 & \cellcolor[rgb]{0.99, 0.99, 0.99} 2.75 & \cellcolor[rgb]{1.00, 1.00, 1.00} 0.00 & \cellcolor[rgb]{0.78, 0.78, 0.78} 33.16 & \cellcolor[rgb]{0.98, 0.98, 0.98} 3.66 & \cellcolor[rgb]{0.96, 0.96, 0.96} 9.39 & \cellcolor[rgb]{0.92, 0.92, 0.92} 14.95 & \cellcolor[rgb]{0.99, 0.99, 0.99} 3.43 & \cellcolor[rgb]{1.00, 1.00, 1.00} 0.17 & \cellcolor[rgb]{0.99, 0.99, 0.99} 1.28 & \cellcolor[rgb]{1.00, 1.00, 1.00} 0.85 & \cellcolor[rgb]{0.98, 0.98, 0.98} 3.91 & \cellcolor[rgb]{0.98, 0.98, 0.98} 4.44\\
    DAC & 8       & \cellcolor[rgb]{0.76, 0.76, 0.76} 35.66 & \cellcolor[rgb]{0.85, 0.85, 0.85} 24.68 & \cellcolor[rgb]{0.98, 0.98, 0.98} 5.00 & \cellcolor[rgb]{1.00, 1.00, 1.00} 0.00 & \cellcolor[rgb]{0.78, 0.78, 0.78} 32.79 & \cellcolor[rgb]{0.98, 0.98, 0.98} 3.95 & \cellcolor[rgb]{0.83, 0.83, 0.83} 27.46 & \cellcolor[rgb]{0.85, 0.85, 0.85} 24.89 & \cellcolor[rgb]{0.77, 0.77, 0.77} 34.57 & \cellcolor[rgb]{0.59, 0.59, 0.59} 76.99 & \cellcolor[rgb]{0.77, 0.77, 0.77} 34.15 & \cellcolor[rgb]{0.95, 0.95, 0.95} 10.21 & \cellcolor[rgb]{0.88, 0.88, 0.88} 20.92 & \cellcolor[rgb]{0.97, 0.97, 0.97} 5.62\\  
 OGG-Opus & 16  & \cellcolor[rgb]{0.59, 0.59, 0.59} 53.44 & \cellcolor[rgb]{0.69, 0.69, 0.69} 41.55 & \cellcolor[rgb]{0.77, 0.77, 0.77} 34.69 & \cellcolor[rgb]{0.79, 0.79, 0.79} 31.73 & \cellcolor[rgb]{0.75, 0.75, 0.75} 37.07 & \cellcolor[rgb]{0.98, 0.98, 0.98} 5.17 & \cellcolor[rgb]{0.74, 0.74, 0.74} 37.46 & \cellcolor[rgb]{0.75, 0.75, 0.75} 36.51 & \cellcolor[rgb]{0.62, 0.62, 0.62} 47.05 & \cellcolor[rgb]{0.59, 0.59, 0.59} 89.91 & \cellcolor[rgb]{0.65, 0.65, 0.65} 44.88 & \cellcolor[rgb]{0.70, 0.70, 0.70} 41.33 & \cellcolor[rgb]{0.70, 0.70, 0.70} 40.67 & \cellcolor[rgb]{0.95, 0.95, 0.95} 10.50\\  
 OGG-Opus & 32   & \cellcolor[rgb]{0.88, 0.88, 0.88} 21.81 & \cellcolor[rgb]{0.96, 0.96, 0.96} 9.16 & \cellcolor[rgb]{0.93, 0.93, 0.93} 14.55 & \cellcolor[rgb]{1.00, 1.00, 1.00} 0.72 & \cellcolor[rgb]{0.77, 0.77, 0.77} 34.30 & \cellcolor[rgb]{0.98, 0.98, 0.98} 4.01 & \cellcolor[rgb]{0.89, 0.89, 0.89} 19.27 & \cellcolor[rgb]{0.93, 0.93, 0.93} 14.76 & \cellcolor[rgb]{0.91, 0.91, 0.91} 17.10 & \cellcolor[rgb]{0.98, 0.98, 0.98} 4.03 & \cellcolor[rgb]{0.88, 0.88, 0.88} 20.53 & \cellcolor[rgb]{0.97, 0.97, 0.97} 6.72 & \cellcolor[rgb]{0.94, 0.94, 0.94} 13.64 & \cellcolor[rgb]{0.97, 0.97, 0.97} 6.59\\ 
 OGG-Opus& 64   & \cellcolor[rgb]{0.98, 0.98, 0.98} 4.57 & \cellcolor[rgb]{1.00, 1.00, 1.00} 0.45 & \cellcolor[rgb]{0.98, 0.98, 0.98} 5.31 & \cellcolor[rgb]{1.00, 1.00, 1.00} 0.00 & \cellcolor[rgb]{0.78, 0.78, 0.78} 33.12 & \cellcolor[rgb]{0.98, 0.98, 0.98} 3.64 & \cellcolor[rgb]{0.95, 0.95, 0.95} 10.90 & \cellcolor[rgb]{0.93, 0.93, 0.93} 13.93 & \cellcolor[rgb]{0.97, 0.97, 0.97} 5.83 & \cellcolor[rgb]{1.00, 1.00, 1.00} 0.29 & \cellcolor[rgb]{0.99, 0.99, 0.99} 3.45 & \cellcolor[rgb]{1.00, 1.00, 1.00} 0.70 & \cellcolor[rgb]{0.98, 0.98, 0.98} 5.40 & \cellcolor[rgb]{0.98, 0.98, 0.98} 4.78\\
 OGG-Opus& 128  & \cellcolor[rgb]{0.99, 0.99, 0.99} 2.29 & \cellcolor[rgb]{1.00, 1.00, 1.00} 0.21 & \cellcolor[rgb]{0.99, 0.99, 0.99} 3.12 & \cellcolor[rgb]{1.00, 1.00, 1.00} 0.00 & \cellcolor[rgb]{0.78, 0.78, 0.78} 33.24 & \cellcolor[rgb]{0.98, 0.98, 0.98} 3.62 & \cellcolor[rgb]{0.96, 0.96, 0.96} 9.32 & \cellcolor[rgb]{0.93, 0.93, 0.93} 14.58 & \cellcolor[rgb]{0.98, 0.98, 0.98} 3.87 & \cellcolor[rgb]{1.00, 1.00, 1.00} 0.14 & \cellcolor[rgb]{0.99, 0.99, 0.99} 1.61 & \cellcolor[rgb]{1.00, 1.00, 1.00} 0.76 & \cellcolor[rgb]{0.98, 0.98, 0.98} 4.07 & \cellcolor[rgb]{0.98, 0.98, 0.98} 4.44\\  
    MP3& 16     & \cellcolor[rgb]{0.59, 0.59, 0.59} 49.80 & \cellcolor[rgb]{0.61, 0.61, 0.61} 48.42 & \cellcolor[rgb]{0.80, 0.80, 0.80} 30.70 & \cellcolor[rgb]{0.97, 0.97, 0.97} 6.82 & \cellcolor[rgb]{0.77, 0.77, 0.77} 34.46 & \cellcolor[rgb]{0.98, 0.98, 0.98} 4.82 & \cellcolor[rgb]{0.75, 0.75, 0.75} 36.28 & \cellcolor[rgb]{0.59, 0.59, 0.59} 78.87 & \cellcolor[rgb]{0.61, 0.61, 0.61} 48.63 & \cellcolor[rgb]{0.59, 0.59, 0.59} 79.14 & \cellcolor[rgb]{0.61, 0.61, 0.61} 48.13 & \cellcolor[rgb]{0.59, 0.59, 0.59} 52.76 & \cellcolor[rgb]{0.74, 0.74, 0.74} 37.25 & \cellcolor[rgb]{0.96, 0.96, 0.96} 8.39\\ 
    MP3& 32     & \cellcolor[rgb]{0.79, 0.79, 0.79} 31.98 & \cellcolor[rgb]{0.95, 0.95, 0.95} 11.52 & \cellcolor[rgb]{0.94, 0.94, 0.94} 12.16 & \cellcolor[rgb]{1.00, 1.00, 1.00} 0.52 & \cellcolor[rgb]{0.77, 0.77, 0.77} 33.62 & \cellcolor[rgb]{0.98, 0.98, 0.98} 4.07 & \cellcolor[rgb]{0.88, 0.88, 0.88} 21.71 & \cellcolor[rgb]{0.92, 0.92, 0.92} 16.13 & \cellcolor[rgb]{0.84, 0.84, 0.84} 26.32 & \cellcolor[rgb]{0.82, 0.82, 0.82} 28.78 & \cellcolor[rgb]{0.84, 0.84, 0.84} 26.24 & \cellcolor[rgb]{0.97, 0.97, 0.97} 6.57 & \cellcolor[rgb]{0.89, 0.89, 0.89} 19.16 & \cellcolor[rgb]{0.97, 0.97, 0.97} 6.00\\
    MP3& 64    & \cellcolor[rgb]{0.97, 0.97, 0.97} 6.28 & \cellcolor[rgb]{1.00, 1.00, 1.00} 0.79 & \cellcolor[rgb]{0.98, 0.98, 0.98} 4.80 & \cellcolor[rgb]{1.00, 1.00, 1.00} 0.04 & \cellcolor[rgb]{0.78, 0.78, 0.78} 33.33 & \cellcolor[rgb]{0.98, 0.98, 0.98} 3.91 & \cellcolor[rgb]{0.94, 0.94, 0.94} 12.38 & \cellcolor[rgb]{0.93, 0.93, 0.93} 14.55 & \cellcolor[rgb]{0.97, 0.97, 0.97} 7.38 & \cellcolor[rgb]{1.00, 1.00, 1.00} 0.33 & \cellcolor[rgb]{0.98, 0.98, 0.98} 5.15 & \cellcolor[rgb]{1.00, 1.00, 1.00} 0.93 & \cellcolor[rgb]{0.97, 0.97, 0.97} 6.76 & \cellcolor[rgb]{0.98, 0.98, 0.98} 4.96\\ 
   MP3& 128     & \cellcolor[rgb]{0.99, 0.99, 0.99} 1.88 & \cellcolor[rgb]{1.00, 1.00, 1.00} 0.29 & \cellcolor[rgb]{0.99, 0.99, 0.99} 2.85 & \cellcolor[rgb]{1.00, 1.00, 1.00} 0.00 & \cellcolor[rgb]{0.78, 0.78, 0.78} 33.18 & \cellcolor[rgb]{0.98, 0.98, 0.98} 3.89 & \cellcolor[rgb]{0.96, 0.96, 0.96} 9.53 & \cellcolor[rgb]{0.92, 0.92, 0.92} 14.84 & \cellcolor[rgb]{0.98, 0.98, 0.98} 3.56 & \cellcolor[rgb]{1.00, 1.00, 1.00} 0.19 & \cellcolor[rgb]{0.99, 0.99, 0.99} 1.34 & \cellcolor[rgb]{1.00, 1.00, 1.00} 0.89 & \cellcolor[rgb]{0.98, 0.98, 0.98} 4.16 & \cellcolor[rgb]{0.98, 0.98, 0.98} 4.78\\  
OGG-Vorbis& qscale=1   & \cellcolor[rgb]{0.63, 0.63, 0.63} 46.56 & \cellcolor[rgb]{0.83, 0.83, 0.83} 27.08 & \cellcolor[rgb]{0.76, 0.76, 0.76} 35.23 & \cellcolor[rgb]{0.81, 0.81, 0.81} 30.23 & \cellcolor[rgb]{0.77, 0.77, 0.77} 34.75 & \cellcolor[rgb]{0.97, 0.97, 0.97} 5.81 & \cellcolor[rgb]{0.75, 0.75, 0.75} 37.03 & \cellcolor[rgb]{0.88, 0.88, 0.88} 21.73 & \cellcolor[rgb]{0.62, 0.62, 0.62} 47.55 & \cellcolor[rgb]{0.59, 0.59, 0.59} 55.22 & \cellcolor[rgb]{0.63, 0.63, 0.63} 46.83 & \cellcolor[rgb]{0.79, 0.79, 0.79} 32.21 & \cellcolor[rgb]{0.76, 0.76, 0.76} 35.93 & \cellcolor[rgb]{0.94, 0.94, 0.94} 12.07\\ 
OGG-Vorbis& qscale=2   & \cellcolor[rgb]{0.69, 0.69, 0.69} 42.17 & \cellcolor[rgb]{0.91, 0.91, 0.91} 17.37 & \cellcolor[rgb]{0.87, 0.87, 0.87} 21.94 & \cellcolor[rgb]{0.93, 0.93, 0.93} 14.39 & \cellcolor[rgb]{0.78, 0.78, 0.78} 33.18 & \cellcolor[rgb]{0.98, 0.98, 0.98} 4.20 & \cellcolor[rgb]{0.85, 0.85, 0.85} 25.45 & \cellcolor[rgb]{0.92, 0.92, 0.92} 14.99 & \cellcolor[rgb]{0.70, 0.70, 0.70} 41.08 & \cellcolor[rgb]{0.59, 0.59, 0.59} 57.72 & \cellcolor[rgb]{0.69, 0.69, 0.69} 41.74 & \cellcolor[rgb]{0.92, 0.92, 0.92} 15.09 & \cellcolor[rgb]{0.82, 0.82, 0.82} 29.07 & \cellcolor[rgb]{0.97, 0.97, 0.97} 6.08\\ 
OGG-Vorbis& qscale=3  & \cellcolor[rgb]{0.73, 0.73, 0.73} 38.10 & \cellcolor[rgb]{0.96, 0.96, 0.96} 9.59 & \cellcolor[rgb]{0.92, 0.92, 0.92} 15.79 & \cellcolor[rgb]{0.96, 0.96, 0.96} 9.24 & \cellcolor[rgb]{0.78, 0.78, 0.78} 33.41 & \cellcolor[rgb]{0.98, 0.98, 0.98} 3.76 & \cellcolor[rgb]{0.88, 0.88, 0.88} 21.07 & \cellcolor[rgb]{0.94, 0.94, 0.94} 13.33 & \cellcolor[rgb]{0.76, 0.76, 0.76} 35.54 & \cellcolor[rgb]{0.85, 0.85, 0.85} 24.87 & \cellcolor[rgb]{0.77, 0.77, 0.77} 34.34 & \cellcolor[rgb]{0.97, 0.97, 0.97} 5.75 & \cellcolor[rgb]{0.86, 0.86, 0.86} 23.61 & \cellcolor[rgb]{0.98, 0.98, 0.98} 4.98\\ 
\midrule
         pooled   &    -   & \cellcolor[rgb]{0.82, 0.82, 0.82} 28.57 & \cellcolor[rgb]{0.92, 0.92, 0.92} 16.38 & \cellcolor[rgb]{0.90, 0.90, 0.90} 19.07 & \cellcolor[rgb]{0.96, 0.96, 0.96} 9.12 & \cellcolor[rgb]{0.77, 0.77, 0.77} 33.87 & \cellcolor[rgb]{0.98, 0.98, 0.98} 4.20 & \cellcolor[rgb]{0.88, 0.88, 0.88} 21.53 & \cellcolor[rgb]{0.90, 0.90, 0.90} 17.81 & \cellcolor[rgb]{0.85, 0.85, 0.85} 24.76 & \cellcolor[rgb]{0.79, 0.79, 0.79} 31.79 & \cellcolor[rgb]{0.84, 0.84, 0.84} 26.35 & \cellcolor[rgb]{0.93, 0.93, 0.93} 13.81 & \cellcolor[rgb]{0.88, 0.88, 0.88} 20.76 & \cellcolor[rgb]{0.97, 0.97, 0.97} 6.43\\ 
      pooled w/o DAC &    -   & \cellcolor[rgb]{0.83, 0.83, 0.83} 27.95 & \cellcolor[rgb]{0.93, 0.93, 0.93} 14.43 & \cellcolor[rgb]{0.89, 0.89, 0.89} 20.19 & \cellcolor[rgb]{0.95, 0.95, 0.95} 9.79 & \cellcolor[rgb]{0.77, 0.77, 0.77} 33.98 & \cellcolor[rgb]{0.98, 0.98, 0.98} 4.22 & \cellcolor[rgb]{0.88, 0.88, 0.88} 20.99 & \cellcolor[rgb]{0.91, 0.91, 0.91} 17.19 & \cellcolor[rgb]{0.86, 0.86, 0.86} 24.06 & \cellcolor[rgb]{0.83, 0.83, 0.83} 28.03 & \cellcolor[rgb]{0.85, 0.85, 0.85} 24.74 & \cellcolor[rgb]{0.93, 0.93, 0.93} 13.96 & \cellcolor[rgb]{0.88, 0.88, 0.88} 20.60 & \cellcolor[rgb]{0.97, 0.97, 0.97} 6.51\\
\bottomrule
\end{tabular}
}
}
\end{table*}

The training configuration is based on the collaborative watermarking scheme proposed in \cite{juvela2024-collaborative-watermarking} and the standard HiFi-GAN training recipe \cite{kong2020-hifi-gan}. The main differences in this work are: 1) all models are now fine-tuned instead of trained from random initialization, 2) the detector model is always AASIST, and 3) the dataset is now LibriTTS-R \cite{koizumi23-libriTTS-R} instead of VCTK \cite{Yamagishi2019-VCTK}.

% Random padding or cropping to 65,536 samples
% Collaborator gradient flow to generator is allowed
% Observer gradients are detached
% Each experiment runs on a single NVIDIA A100 GPU

% reset optimizers and learning rate schedulers

For minibatch training, we cropped or padded the speech utterances to 65,536 samples length and used a batch size of 16. 
All training uses the AdamW optimizer with $\beta_1$ = 0.8, $\beta_2$ = 0.99 and 
exponential learning rate decay with decay factor 0.999. Optimizer and learning rate scheduler states are reset at the start of fine-tuning.
Observer settings use the original HiFi-GAN initial learning rate of 2e-4, while for collaborator settings the initial learning rate is reduced to 2e-5. 
We use different learning rates because
Collaborators tend to converge very fast but suffer from training instability at the higher learning rate. In contrast, the observers do not converge fast enough at the lower learning rate. Additionally, we clip the gradient norm in both detector models for gradient values exceeding unit magnitude over the time dimension.
All experiment configurations are fine-tuned for 20 epochs, amounting to 40k parameter updates. We used NVIDIA A100 GPUs for training, and each model was trained on a single GPU.

\section{Results}

\subsection{Detection equal error rates}

Equal error rates (EERs) are listed in Table
\ref{tab:eer}. Table columns represent various codec augmentation configurations, while rows correspond to evaluation conditions. Columns alternate between Observer and Collaborator training (labeled `o' and `c', respectively) for otherwise matching configurations. System augmented with DAC has only seen DAC during training, while the systems with MP3/Opus augmentation choose one of the codecs randomly. The leftmost column used no codec augmentation and corresponds to fine-tuning the detector and HiFi-GAN models. The rightmost column for `MP3/Opus all' has seen both MP3 and Opus at all four bitrates.
%
%Detection was done in randomly cropped 65,536 sample windows, which matches the training configuration. 
Detection was conducted by feeding the entire waveform to the detector. 
Evaluation EERs are averaged over five repetitions over the test set.

\subsection{Listening test}

We conducted a mean opinion score (MOS) test to quantify the perceptual degradation related to codec augmentation of collaborative detector training at various bitrates. Baseline systems include natural speech, a pre-trained HiFi-GAN V1 Universal model, and a fine-tuned HiFi-GAN matching the number of fine-tuning iterations used in Collaborator and Observer training. 
The test stimuli consist of 50 utterances of duration between 3 and 10 seconds randomly selected from the test set. Each utterance was rendered with all test systems, no codecs were applied on the listening test samples and each sample was loudness-normalized to -24 LUFS.

\begin{figure}[htb]
    \centering
    \includegraphics[width=0.95\linewidth]{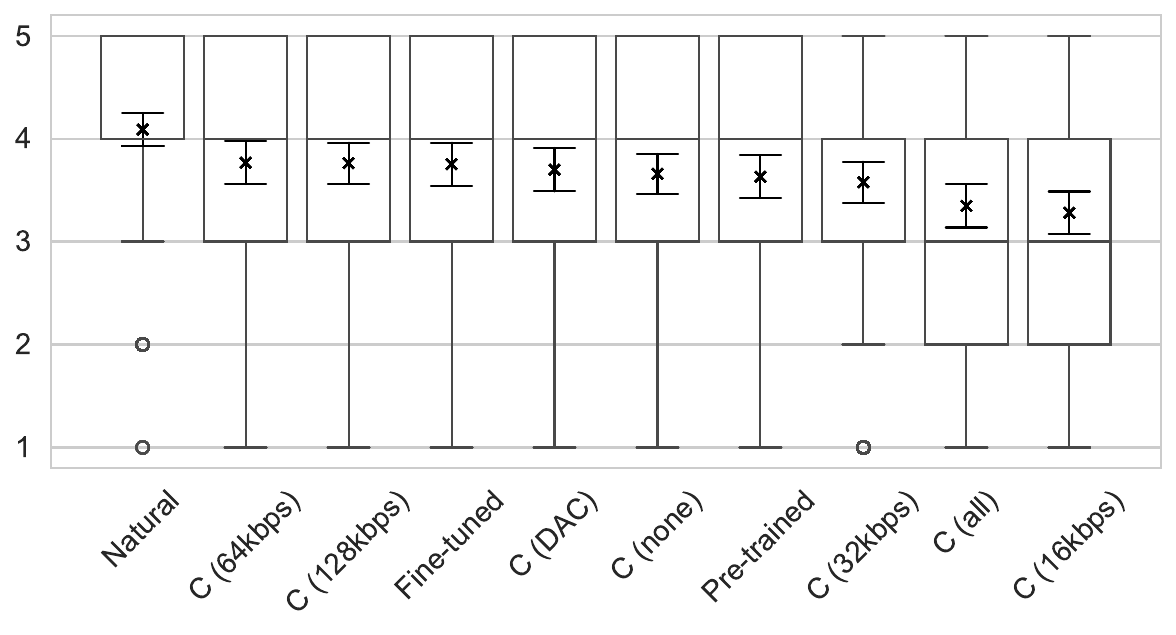}
    \caption{Mean opinion scores (MOS) for speech naturalness. `C' denotes collaborative training, with codec augmentation type in parentheses.}
    \label{fig:mos-scores}
\end{figure}

% Tricks for boosting MOS scores:
% more granular scale: half points possible
% instruct listeners to rate natural as 5
% filter out subjects who rate natural speech poorly
Listening was conducted on the Prolific\footnote{\url{https://www.prolific.com/}} crowd-sourcing platform.  Listeners were asked to rate the naturalness of speech samples on a five-level Likert scale from 1 (Bad) to 5 (Excellent) and were requested to wear headphones during the test. The test samples were randomly divided to assignment batches of size 125, and each batch was rated by 5 listeners. In total, the evaluation consisted of 20 listeners giving 2500 individual ratings. 
All subjects rated natural samples above 3 and no ratings were filtered out.
Fig.~\ref{fig:mos-scores} shows a boxplot of the ratings overlaid with 
mean ratings and t-statistic based 95\% confidence intervals adjusted for multiple comparisons using the Dunn-Bonferroni correction.

\section{Discussion and Conclusion}

Comparing the detection performance in Table \ref{tab:eer} and listening test results in Fig. \ref{fig:mos-scores} lets us draw some conclusions about the usefulness of codec augmentation in collaborative and passive detection. Note that all Observer settings correspond to fine-tuned HiFi-GAN in the listening test, since the detector has no effect on the generator.
First, at high bitrates (64kbps and 128kpbs) the baseline systems without augmentation perform quite well, and the corresponding augmented systems have very similar performance both in EER and MOS.
%
%Collaborative training with Codec-STE is useful for trading off perceptual quality to robustness against low rate codecs
Second, we can observe that Collaborative training can be paired with lower bitrate codec augmentation to trade off detection robustness to some perceptual degradation. Collaborative training at 16\,kbit augmentation leads to the most robust detection, but also the worst perceptual quality. 
Pooling all bitrates together gives a similar result. This is somewhat expected, since the generator has no knowledge of the channel and must prepare for worse case.

Finally, as a perhaps surprisingly positive result, augmentation using only the DAC neural audio codec transfers quite well to evaluation on traditional audio codecs. This is especially prominent in the collaborative setting, and DAC augmentation only causes minor perceptual degradation.
Future work includes extending the collaborative watermarking method to a wider range of generative systems and establishing a comprehensive robustness evaluation framework.

\bibliographystyle{IEEEtran}
\bibliography{references}

\end{document}